# Is the second harmonic method applicable for thin films mechanical properties characterization by nanoindentation?

G. Guillonneau[a,b*], G. Kermouche[c], J. Teisseire[d], E. Barthel[d], S. Bec[a], J.-L. Loubet[a]

[a]Ecole Centrale de Lyon, Laboratoire de Tribologie et Dynamique des Systèmes, UMR 5513 CNRS/ECL/ENISE, 69134 Ecully, France

[b]EMPA, Swiss Federal Laboratories for Materials Science and Technology, Laboratory for Mechanics of Materials and Nanostructures, Feuerwerkerstrasse 39, CH 3602 Thun, Switzerland

[c]Ecole des Mines de Saint-Etienne, Centre SMS, Laboratoire LGF UMR5307, 158 Cours Fauriel 42023 Saint-Etienne

[d]Laboratoire « Surface du Verre et Interfaces », CNRS/Saint-Gobain UMR 125, 39, quai Lucien Lefranc, BP 135, F-93303 Aubervilliers Cedex

[*]Corresponding author: gaylord.guillonneau@empa.ch

# Is the second harmonic method applicable for thin films mechanical properties characterization by nanoindentation?

Abstract

The second harmonic method is a dynamic indentation technique independent of the direct indentation depth measurement. It can be used to determine near-surface mechanical properties of bulk materials more precisely than classical dynamic nano-indentation. In this paper, the second harmonic method is extended to the measurement of the mechanical properties of thin PMMA layers deposited onto silicon wafers. It is shown that this new technique gives precise results at small depths (less than 100nm), even for films with a thickness lower than 500nm, which was not possible to achieve with the classical CSM method. However, experimental and numerical results obtained both with classical nanoindentation and second harmonic methods differ at high indentation depth. Using FE simulations and AFM measurements, it is shown that the contact depth calculation with classical models can explain this difference.

## 1. Introduction

### 1.1. About nanoindentation measurement

Because the nanoindentation is a non-destructive technique which measures the mechanical properties at the nanometer scale, it is particularly adapted to probe surface properties of very thin films. The principal properties measured with this apparatus are the reduced elastic modulus $E'^*$ and the hardness $H$ of the tested material [1,2]:

$$H = \frac{P}{A_c} \tag{1}$$

$$\frac{1}{E'^*} = \frac{1}{E_c'^*} - \frac{1}{E_i'^*} \tag{2}$$

With:

$$E_c^{'*} = \frac{S}{2}\sqrt{\frac{\pi}{A_c}}$$  (3)

Where $P$ is the load, $A_c$ is the projected contact area, $S$ is the harmonic contact stiffness, $E_i^{'*}$ is the reduced elastic modulus of the tip, and $E_c^{'*}$ is the reduced contact modulus between the sample and the tip. The contact area $A_c$ is directly linked to the contact depth $h_c$ by the following geometrical equation (for a perfect pyramidal indenter) [3]:

$$A_c = \pi \tan^2(\theta) h_c^2$$  (4)

Where $\theta$ is the equivalent semi-angle of the tip. This expression can be changed in order to take into account the non-perfect geometry of the indenter [3]. The sample stiffness $S$ can be measured by fitting the unloading curve [3], permitting to obtain $E$ and $H$, at maximum load, or with the dynamic CSM (Continuous Stiffness Measurement) technique. The CSM method consists of adding a small oscillation to the load signal in order to measure the dynamic contact stiffness as a function of the indentation depth [4,5]. This dynamic technique is particularly used for thin films characterization, because the mechanical properties of the layer can be measured as a function of the indentation depth with only one test, which is impossible with the quasi-static method.

*1.2. About indentation of thin films and related difficulties*

Thin film indentation differs from bulk material characterization, and some difficulties have to be considered. First, the substrate properties have to be taken into account. Indeed, during thin film indentation, the measured properties are a combination of the film properties and the substrate properties. To account for the substrate effect, several literature models, more or less complex, are proposed. Nix proposes a simple model to determine the film modulus, dependent on an adjustable parameter [6]. Rar et al. developed a more complex model, requiring the knowledge of the Poisson ratio of the tested materials [7,8]. Perriot and Barthel proposed another elastic model to take into account the substrate effect. This model can be expressed by the following equation [9]:

$$E^{'*} = E_f^{'*} + \frac{E_s^{'*} - E_f^{'*}}{1 + \left(\frac{tx_0}{\tan(\theta)h_c}\right)^n} \qquad (5)$$

Where $E_f^{'*}$ is the film modulus, $E_s^{'*}$ the substrate modulus, $t$ the film thickness, and $n$ and $x_0$ are coefficients dependent of the ratio $E_s^{'*}/E_f^{'*}$ [9]. This model gives precise results for a large range of $E_s^{'*}/E_f^{'*}$ ratios. A simpler expression, proposed by Bec et al., can be used to link the apparent reduced elastic modulus $E^{'*}$ to the reduced modulus of the film and the substrate [10,11]:

$$\frac{1}{E^{'*}} = \left(1 + \frac{2}{\pi}\frac{t}{\tan(\theta)h_c}\right)^{-1}\left(\frac{2}{\pi E_f^{'*}}\frac{t}{\tan(\theta)h_c} + \frac{1}{E_s^{'*}}\right) \qquad (6)$$

This model gives good results for a large range of materials and is not dependent of adjustable parameters.

The second difficulty, common to bulk materials, comes from the contact depth calculation. The contact depth value cannot be directly measured, but can be calculated with models (in order to model pile-up or sink-in phenomena) using the displacement measurement $h$. The most popular is the Oliver and Pharr model which links the contact depth $h_c$ to the indentation depth $h$ by the following equation [3,5]:

$$h_c = h - \varepsilon \frac{P}{S} \qquad (7)$$

Where $\varepsilon$ is a parameter which depends on the tip geometry ($\varepsilon=0.75$ for a Berkovich indenter). Another expression proposed by Loubet et al., can be used [12,13]:

$$h_c = \alpha \left( h - \frac{P}{S} + h_0 \right) \qquad (8)$$

Where $\alpha=1.2$ for a Berkovivh indenter and $h_0$ is a coefficient calculated to take into account the tip defect [13,14].

The third difficulty comes from the uncertainties linked to the displacement measurement. As explained in a previous paper, the displacement measurement is influenced by uncertainties like contact point detection, tip defect, thermal drift, or sample roughness [14,15]. These uncertainties are particularly important at very small penetration depths, resulting in less precise mechanical properties measurement at very small indentation depths, which can be problematic when the film thickness is lower than 100 nm.

### 1.3. What could we expect from the use of the second harmonic method?

The second harmonic method can be used to measure the mechanical properties of bulk materials at indentation depths lower than 100 nm with better accuracy than classical CSM method because the technique is independent of the direct displacement measurement [16]. Consequently, the method is expected to give precise values of mechanical properties at small indentation depth for thin films. In this paper, an application of the second harmonic method on thin PMMA layers deposited on silicon wafer is presented. In the following part, the expressions used to calculate thin films properties with the second harmonic method are detailed. In the third part, the experimental tests, the samples and the apparatus are described. The results are shown and discussed in the fourth part, and we discuss the difference between results at high indentation depths in the last part.

## 2. Second harmonic method: substrate effect

In a previous paper, it was shown how to measure the mechanical properties of bulk materials with the second harmonic method [16]. In the case of thin film indentation the same approach can be used, but the calculation of *dS/dh* (equation (11) of [16]) has to consider the reduced contact modulus variation as a function of the indentation depth, because of the substrate's elasticity influence. The Sneddon relation links the stiffness to the reduced contact modulus and the contact depth by the following equation:

$$S = 2 E_c^{'*} \tan(\theta) h_c \qquad (9)$$

Considering the variation of $E_c^{'*}$ and $h_c$ with the indentation depth, and using Equations (3) and (9), the derivative of Equation (9) with respect to the displacement gives:

$$\frac{dS}{dh} = 2 E_c^{'*} \tan(\theta) \frac{dh_c}{dh} - S E_c^{'*} \frac{d(1/E^{'*})}{dh} \qquad (10)$$

The expression depends on the indenter geometry, the reduced contact modulus, $dh_c/dh$, and $d(1/E^{'*})/dh$. The derivative of the contact depth with respect to the indentation depth can be determined by a simple local derivative of the $h_c$-$h$ curve. However, the term $d(1/E^{'*})/dh$ is not known. We propose the use of a literature model to calculate this expression. Because of its simplicity, and also because it does not depend on adjustable parameters, the Bec et al. model was used to calculate $d(1/E^{'*})/dh$ [10,11]. After derivation of Equation (6), the term $d(1/E^{'*})/dh$ can be obtained:

$$\frac{d(1/E^{'*})}{dh} = \frac{\frac{2t}{\pi \tan(\theta)h_c^2}\left(\frac{1}{E_s^{'*}} - \frac{1}{E_f^{'*}}\right)}{1 + \left(\frac{2t}{\pi \tan(\theta)h_c}\right)^2}\frac{dh_c}{dh} \quad (11)$$

In this expression, all terms can be measured or calculated. Finally, with Equations (10) and (11), the expressions of $E_c^{'*}$ and $H$ can be obtained:

$$E_c^{'*} = \frac{\frac{dS}{dh}}{2\tan(\theta)\frac{dh_c}{dh} - S\frac{d(1/E^{'*})}{dh}} \quad (12)$$

$$H = \frac{4P}{\pi S^2}E_c^{'*2} \quad (13)$$

And $E^{'*}$ is calculated using Equation (2).

## 3. Experimental details

### 3.1. Tested samples

Experimental tests were performed on poly(methyl methacrylate) (PMMA) (Sigma-Aldrich, St Quentin Fallavier) [Mw=15 kg/mol, entanglement molecular weight Me = 10 kg/mol]. Thin layers were deposited by spin-coating onto clean silicon wafer Si(100) (Neyco). Five samples were tested with different PMMA film thicknesses *t,* adjusted by dilution in methyl-isobutylketone (185 nm, 348 nm, 631 +/-3 nm, 982 +/-3 nm and 1914 +/-3 nm, thicknesses measured by AFM). The residual stresses and residual solvent in the film were minimized by annealing at 170 °C for 20 min and cooling down to room temperature in 4 h. The samples were cut, then glued with cyanoacrylate on an aluminum sample holder, and cleaned with ethanol using a standard paper towel.

*3.2. Apparatus*

A SA2® Nanoindenter equipped with a DCM head was used in the experiments. The resolutions in force and displacement are 1nN and 0.2pm respectively. The apparatus and the second harmonic measurement method were described in more details in a previous paper [14]. A sharp diamond Berkovich tip was used in the experiments ($h_0$=5 nm, calculated using the Loubet et al. method [13]). For each sample, 10 tests were done to obtain a mean value of the properties. The maximum load was 10 mN. Because the PMMA is a material whose properties are time dependent, a constant strain rate was applied with $\dot{P}/P$=0.01s$^{-1}$ [17,18]. The CSM technique was applied with an oscillation amplitude varying between 2 and 4 nm following the indentation depth value [19], and a frequency of 31 Hz.

Mechanical properties are computed using both the classical CSM method and second harmonic method and are plotted as a function of the ratio a/t (contact radius/layer thickness). Results for depths lower than 25 nm are not plotted in the paper, the tip defect being too much important at lower depths to measure precise mechanical properties. The theoretical values of the reduced elastic modulus used in Equation (11) are $E_f^{'*}$ = 5 GPa for PMMA and $E_s^{'*}$ =180 GPa for silicon [10,14]. The contact depth used in Equation (2) is calculated with the Loubet et al. model [12,13].

## 4. Results

*4.1. Hardness*

From Figures 1 and 2, the hardness of PMMA layers is approximately constant for small *a/t* values (*a* is the contact radius) with the classical CSM method and second harmonic method, and is in good agreement with the PMMA hardness [20]. With the second harmonic method, the hardness curves are superposed at very small *a/t* ratio. It is not the case with the classical CSM method, as some scattering in hardness value is observed at low *a/t* ratio. This scattering at low depths is caused by of the uncertainties related to the tip displacement, especially the contact point detection and the tip defect. This confirms that the second harmonic method is particularly adapted to the measurement of mechanical properties at small indentation depths, given that it is not dependent of the displacement measurement. An increase of the hardness can be observed for CSM method from $a/t = 0.8$, showing the substrate effect on the measurement. The scattered observed in the curves is due to the stiffness measurement which is scattered at this range. With the second harmonic method, the hardness increase begins for smaller $a/t$ ratio ($a/t$=0.2), for samples with thickness between 631 and 1914nm. This difference will be explained in section 5.2.

*4.2. Elastic modulus*

On Figures 3 and 4, the experimental reduced elastic modulus determined with the two techniques is compared to values calculated with the Bec et al. and the Perriot and Barthel models. Good correlation between experimental and numerical results is observed for small $a/t$ ratios in both methods. Furthermore, as obtained on hardness results, the elastic modulus measured with the second harmonic method is more precise than the modulus determined with the classical CSM technique for the same reason than in the previous paragraph. However, for higher $a/t$ ratios, the experimental elastic modulus is higher than calculated values. With the CSM method, the curves differ from $a/t$=1, and with the second harmonic method, the curves differ from a/t=0.4. Moreover, the elastic modulus calculated with the second harmonic method is twice higher than the modulus determined with the CSM method.

A first explanation of this divergence can come from the contact depth calculation. Indeed, with their numerical model, Perriot and Barthel show a strong difference between the real contact depth and $h_c$ calculated with the literature models when the $E_s^{'*}/E_f^{'*}$ ratio is very high or very low [9]. The contact depth is calculated supposing a homogeneous material

(which is not the case for thin films indentation) and Perriot and Barthel explained that this hypothesis is the cause of this divergence.

The CSM method is based on the calculation of $h_c$, and the second harmonic method is based on $h_c$ and $dh_c/dh$ determination. Consequently, if $h_c$ is not correctly calculated, its derivative also can be not correctly calculated. This explains the large difference between experimental and numerical results. In the following paragraph, a Finite Element simulation of the PMMA/Si indentation and AFM measurements are presented in order to further investigate the contact depth calculation.

## 5. Finite Element analysis

### *5.1. Finite Element model*

Calculations were performed with Systus/Sysweld [21,22] using axisymmetric elements and a large displacement / large strain option (updated Lagrangian formulation). Here the commonly-used assumption in which the Berkovich tip can be replaced by a cone of semi angle 70.32° is considered. The Si thickness has been chosen to be one hundred times the maximum contact area. To ensure plastic incompressibility, four node quadrilateral iso-parametric elements with a selective reduced integration scheme are used in the plastically deformed area. The plastic flow is described via a plastic von Mises stress. The loading is achieved by imposing a quasi-static displacement of the indenter.

One important difficulty related to the Finite Element analysis of the indentation of soft thin films over hard substrate is the large mesh distortion of the film at high penetration depth and the necessity to use very small elements at low penetration depth. The only way to bypass this issue is to use an adapted remeshing procedure. For that purpose, we developed an automatic remeshing procedure based on the work of Kermouche et al. in the case of scratch testing of coating/substrate system [23]. The remeshing strategy is based on the definition of several zones in which the mesh density is different. A very fine mesh is defined near the contact. In the other zones, the greater the distance to the fine mesh zone, the lower the mesh density. During a remeshing stage, these zones are calculated starting from the coordinates of the indenter and the parameters of the contact area (contact radius, contact depth). The meshing algorithm is such that the quantity of nodes in contact is maintained constant during the calculation and thus does not depend on the penetration depth of the indenter. The node quantity in the film thickness is computed from the size of the contact area. Indeed, when the thickness of the film is negligible compared to the contact radius, it is not necessary to describe finely what happens in the film. In the opposite case, when the contact radius is lower than the film thickness, an important elements quantity is required, because the deformation of the film may play a very significant role. Therefore, it is possible to accurately determine the contact area and the contact stiffness whatever the indentation depth by superimposing small oscillation amplitude to the imposed displacement in order to simulate the Continuous Stiffness Measurement method.

The diamond and silicon are supposed to follow a linear isotropic elastic behavior, and the PMMA layer is supposed to be elastic- perfectly plastic. Note that the well-known time-dependence of PMMA mechanical properties [18,21,24] is not taken into account to simplify the model and the analysis. This will be a future issue of this work. The elastic modulus of diamond is $E_i^*$ =1150 GPa. The PMMA yield stress is $\sigma^y$ =100 MPa [21,25].

*5.2. Results*

On Figure 5, the composite reduced elastic modulus determined by Finite Element simulation is in good agreement with values obtained by literature models, at both small and high *a/t* ratios. It is satisfying because the modulus was calculated only from the basis of computational data and was not affected by the use of any model to estimate the contact depth.

On Figure 6, the $h_c/h$ ratio determined by Finite Element simulation is compared with values calculated with the Loubet model, the Oliver and Pharr model, and the Perriot and Barthel model [3,9,12,13]. These three models were computed from the contact stiffness, indentation depth and load extracted from the Finite Element analysis. The $h_c/h$ ratio calculated with the Loubet and the Oliver and Pharr models are almost independent of the penetration depth, contrary to the $h_c/h$ ratio calculated from the Finite Element simulation. From $a/t>1.5$, this $h_c/h$ ratio is significantly higher than values obtained with classical models, which can explain why the experimental elastic modulus determined with CSM method is overestimated. Moreover, the $h_c/h$ ratio is higher than 1 from $a/t>1.5$, which means that pile-up occurs around the indenter as shown in Figure 8. This is confirmed by the residual print morphology measured with the help of an Atomic Force Microscope (Figure 9).

Consequently, the contact depth cannot be calculated correctly with the Oliver and Pharr model. It can also be observed that the $h_c/h$ ratio obtained with the Perriot and Barthel model is the lower one. This result is not surprising because this model is based on the assumption that both film and substrate follow a linear isotropic elastic behavior. Note however that its variation follows the same trend of the Finite Element results.

As the second harmonic method depends also on the derivative of the contact depth with respect to the indentation depth, the variation of $dh_c/dh$ versus $a/t$ ratio was plotted in Figure 7. The same trends are observed. The three contact models fail to compute accurately $dh_c/dh$ except for low $a/t$ ratio (typically $a/t<0.4$ from the Loubet et al. model). Consequently the combination of the errors in the computation of $h_c/h$ and $dh_c/dh$ makes the second harmonic method more inaccurate than the CSM method in the case of film/substrate system for high $a/t$ ratio. More precisely, the ratio $a/t$ below which the second harmonic method can be applied is lower than the one corresponding to the CSM method as observed on Figures 3 and 4.

All these results show the importance of the correct calculation of the contact depth for thin film indentation. However, this error in the calculation of $h_c$ and $dh_c/dh$ is not sufficient to totally explain the difference. Another phenomenon linked to the dependence of elastic properties of PMMA to the hydrostatic pressure can be considered [26,27]. A Finite Element simulation of the PMMA/Si indentation with elastic properties of PMMA dependent on the hydrostatic pressure could be a perspective complementary to this work.

## 6. Conclusion

In this paper, the second harmonic method was applied to thin PMMA layers deposited onto silicon wafer. To calculate correctly the mechanical properties, it was necessary to recalculate the derivative of the contact stiffness with respect to the displacement in order to account for the variation of the elastic modulus versus indentation depth. This calculation was performed using the model of Bec et al. [10,11]. Experimental application shows precise properties measured at small indentation depths with both CSM and second harmonic methods, with a better accuracy for second harmonic method at low *a/t* ratio. At high indentation depth, a significant difference between experimental and numerical results was observed. A Finite Element investigation coupled with AFM observation of indentation print evidenced that classical contact models fail to compute accurately both $h_c/h$ ratio and $dh_c/dh$ leading thus to an overestimation of the mechanical properties.

Finally, one important conclusion of this paper is that both the CSM method and the second harmonic method are not accurate enough to measure the mechanical properties of film/substrate system over a large range of *a/t* ratio using standard contact models. With the second harmonic method, it is possible to measure accurately the mechanical properties at low *a/t* values, where the CSM method is subjected to various measurement uncertainties. On the contrary, with the CSM method, it is possible to obtain accurate results over a larger range of *a/t* ratio (until *a/t*=1). The reason why the CSM method fails at larger indentation depth may be linked to the bad estimation of the contact area but it can also be related to the possible hydrostatic pressure dependence of the elastic properties of PMMA when it is confined between the tip and the silicon substrate. This last point will be considered in a future paper.

## 8. Figures

Figure 1. Hardness versus $a$/t ratio (contact radius/film thickness) for PMMA/Si samples, calculated with the CSM method. Each curve corresponds to the mean value from 10 indentation tests.

Figure 2. Hardness versus $a$/t ratio (contact radius/film thickness) for PMMA/Si samples, calculated with the second harmonic method. Each curve corresponds to the mean value from 10 indentation tests.

Figure 3. Composite reduced elastic modulus versus $a/t$ ratio (contact radius/film thickness) for PMMA/Si samples calculated with the CSM method. Each curve corresponds to the mean value from 10 indentation tests. Comparison with modulus calculated with literature models.

Figure 4. Composite reduced elastic modulus versus $a/t$ ratio (contact radius/film thickness) for PMMA/Si samples, calculated with the second harmonic method. Each curve corresponds to the mean value from 10 indentation tests. Comparison with modulus calculated with literature models.

Figure 5. Composite reduced elastic modulus versus $a$/t ratio (contact radius/film thickness) for PMMA/Si sample, determined by Finite Element simulation. Comparison with modulus calculated with literature models.

Figure 6. $h_c/h$ ratio versus $a/t$ ratio (contact radius/film thickness) for PMMA/Si sample obtained by F.E. simulation. Comparison between the value calculated from F.E. simulation and $h_c/h$ calculated with literature models using the simulation data as input.

Figure 7. $dh_c/dh$ versus $a/t$ ratio (contact radius/film thickness) for PMMA/Si sample obtained by F.E. simulation. Comparison between the value obtained with F.E. simulation and $dh_c/dh$ calculated with literature models.

Figure 8. Von Mises stress distribution near the contact during indentation on PMMA layer from EF simulation.

Figure 9. a) AFM image (error mode) of an indentation print on a PMMA/Si layer (t=982 nm). b) Topography profile along the line.

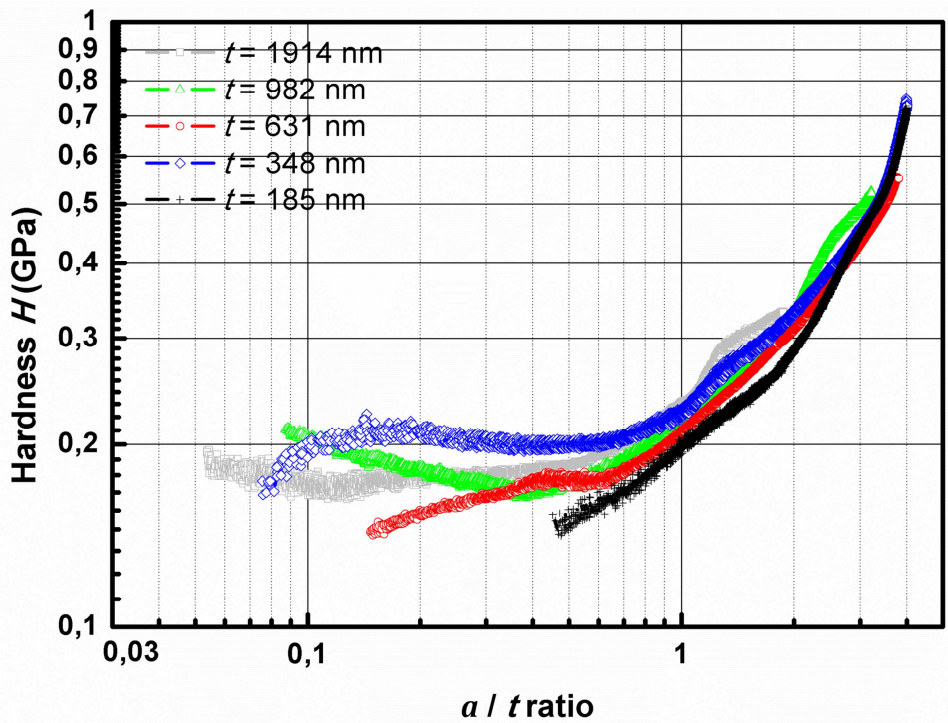

Figure 1. Hardness versus *a*/t ratio (contact radius/film thickness) for PMMA/Si samples, calculated with the CSM method. Each curve corresponds to the mean value from 10 indentation tests.

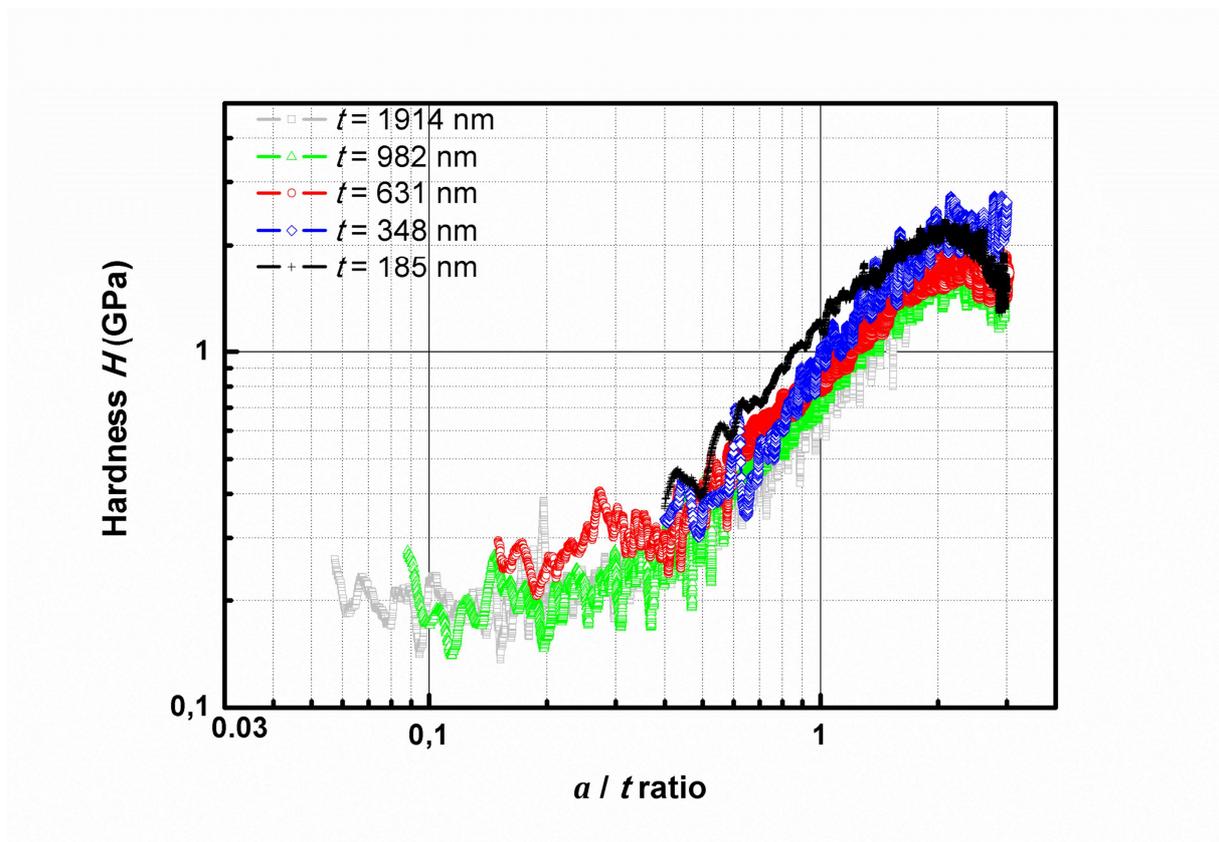

Figure 2. Hardness versus *a*/t ratio (contact radius/film thickness) for PMMA/Si samples, calculated with the second harmonic method. Each curve corresponds to the mean value from 10 indentation tests.

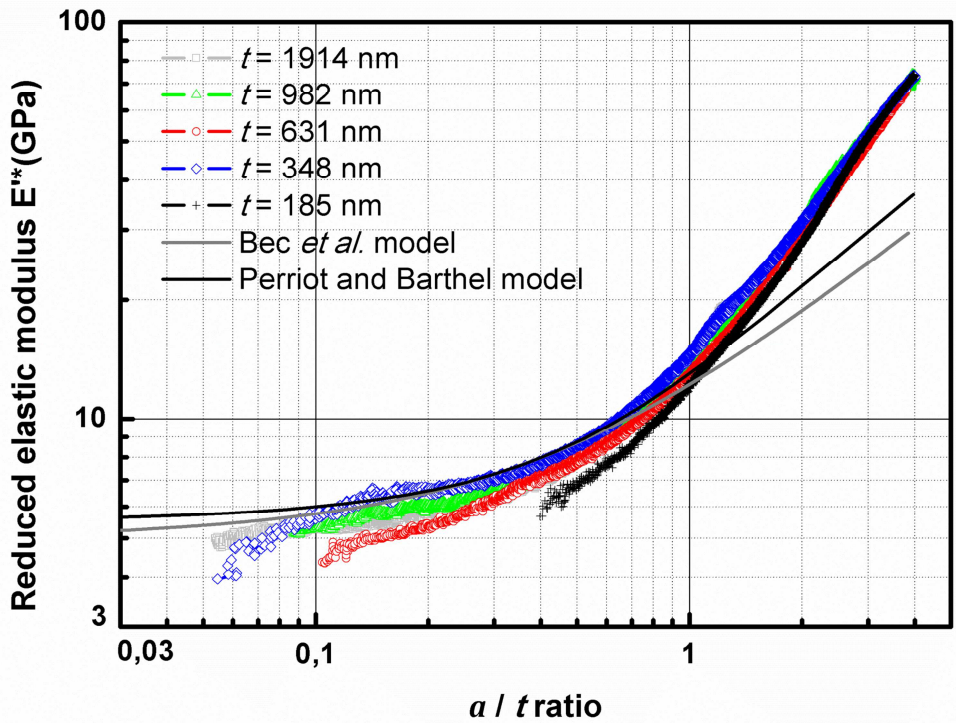

Figure 3. Composite reduced elastic modulus versus *a/t* ratio (contact radius/film thickness) for PMMA/Si samples calculated with the CSM method. Each curve corresponds to the mean value from 10 indentation tests. Comparison with modulus calculated with literature models.

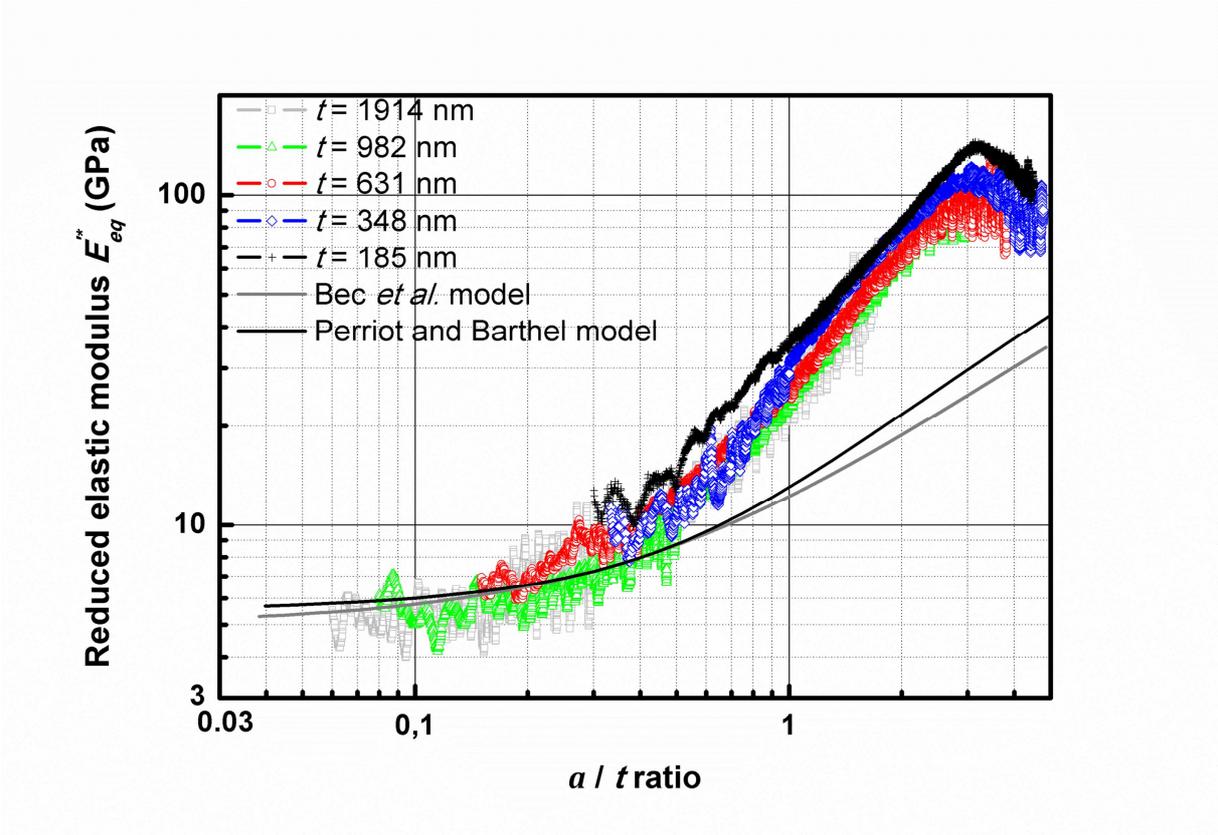

Figure 4. Composite reduced elastic modulus versus *a/t* ratio (contact radius/film thickness) for PMMA/Si samples, calculated with the second harmonic method. Each curve corresponds to the mean value from 10 indentation tests. Comparison with modulus calculated with literature models.

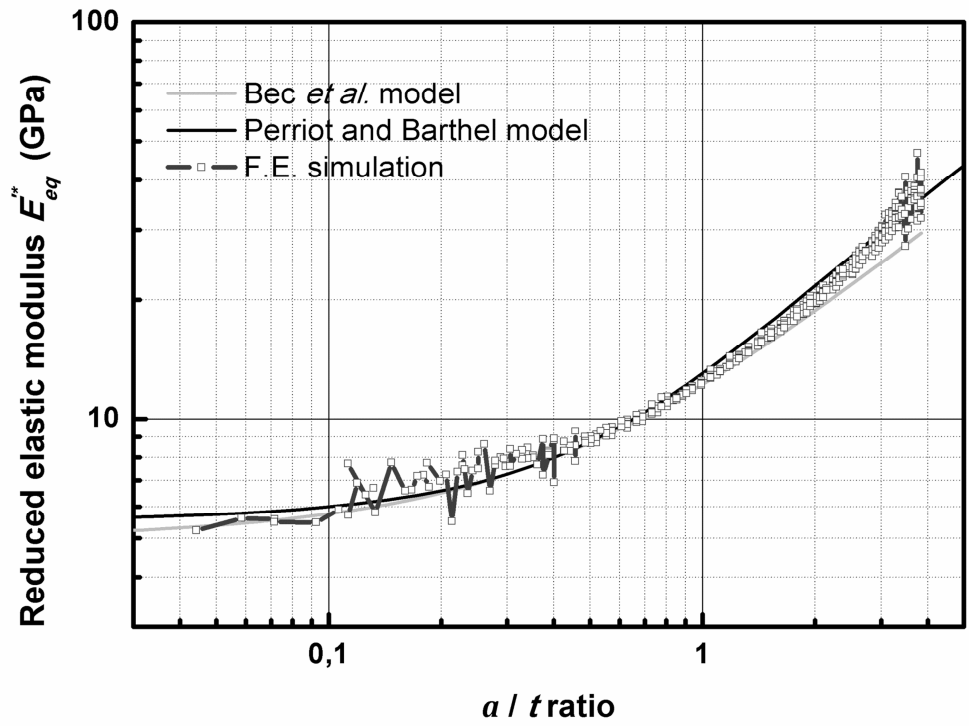

Figure 5. Composite reduced elastic modulus versus *a*/t ratio (contact radius/film thickness) for PMMA/Si sample, determined by Finite Element simulation. Comparison with modulus calculated with literature models.

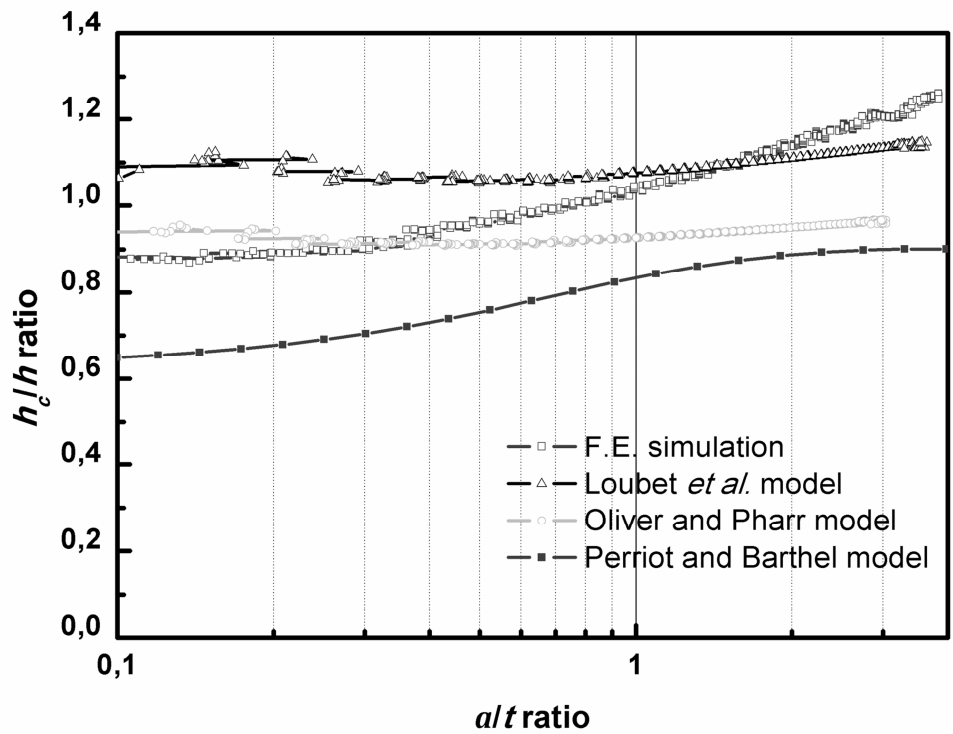

Figure 6. $h_c/h$ ratio versus $a/t$ ratio (contact radius/film thickness) for PMMA/Si sample obtained by F.E. simulation. Comparison between the value calculated from F.E. simulation and $h_c/h$ calculated with literature models using the simulation data as input.

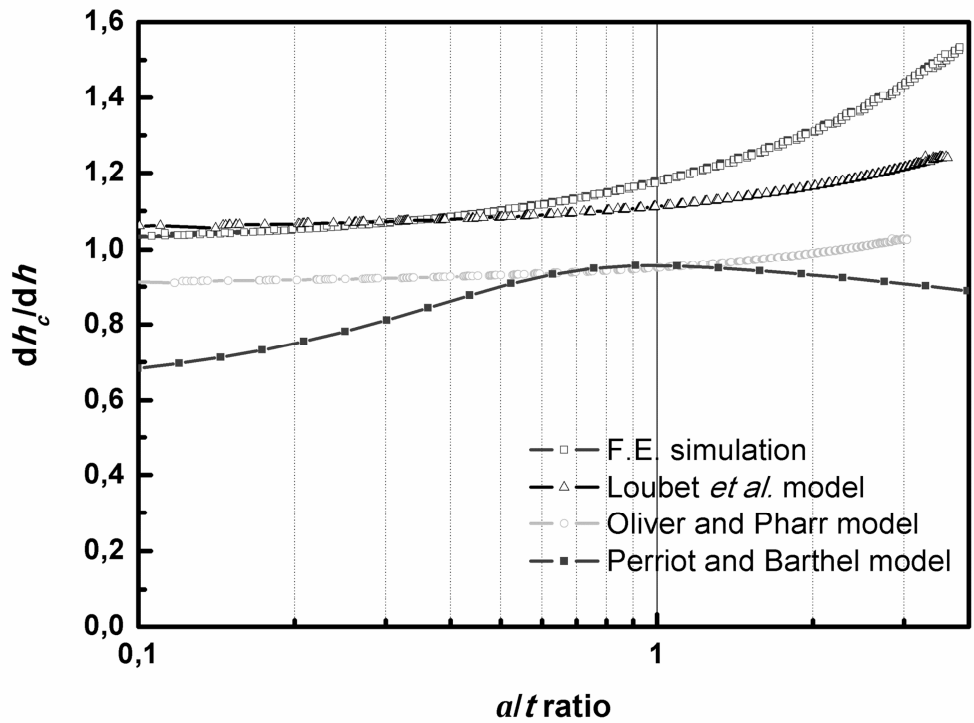

Figure 7. $dh_c/dh$ versus $a/t$ ratio (contact radius/film thickness) for PMMA/Si sample obtained by F.E. simulation. Comparison between the value obtained with F.E. simulation and $dh_c/dh$ calculated with literature models.

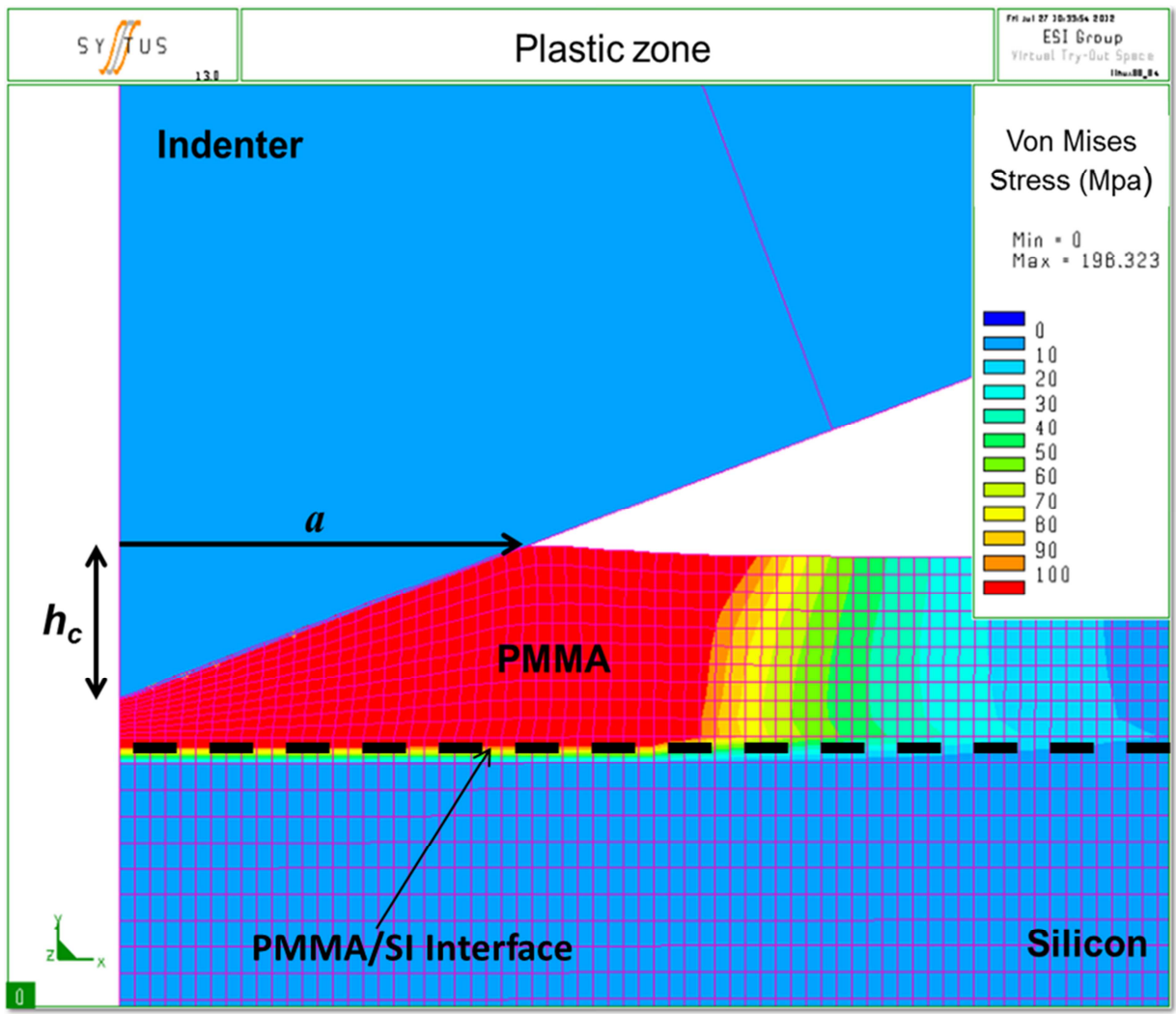

Figure 8. Von Mises stress distribution near the contact during indentation on PMMA layer from EF simulation.

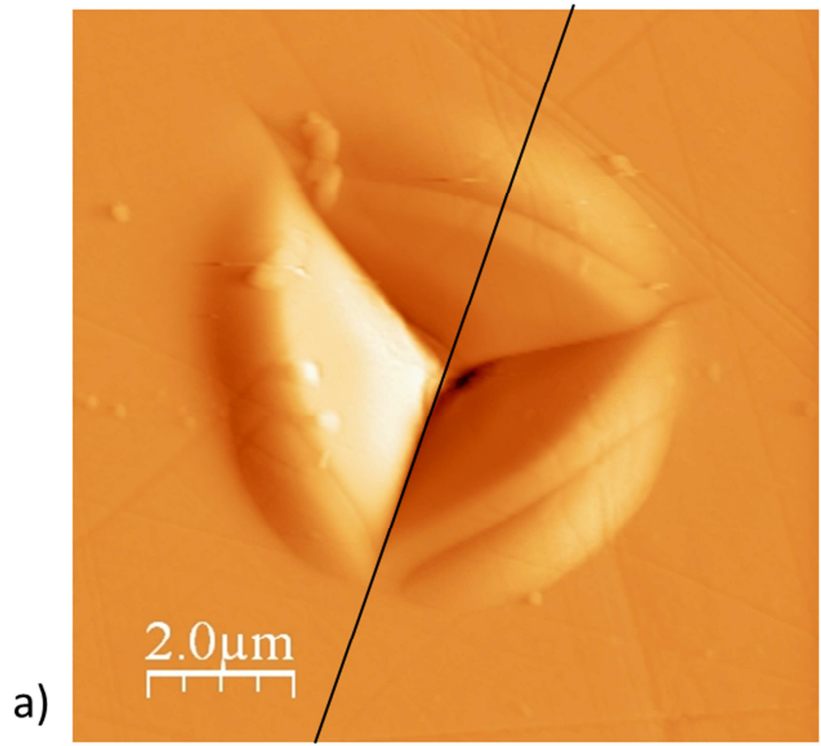

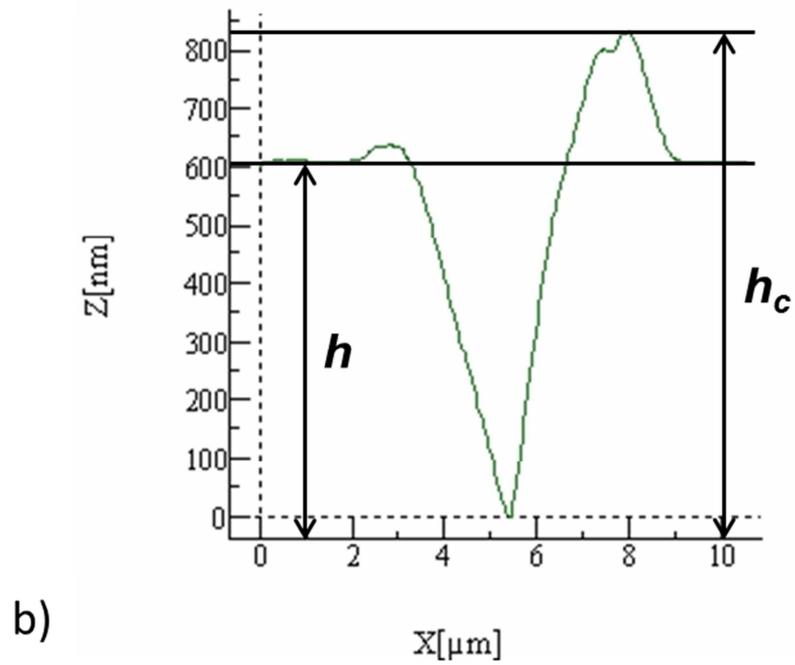

Figure 9. a) AFM image (error mode) of an indentation print on a PMMA/Si layer (t=982 nm). b) Topography profile along the line.